\documentclass[twocolumn,showpacs,preprintnumbers,amsmath,amssymb]{revtex4}
\usepackage{graphicx}
\usepackage{epsfig} 
\usepackage{dcolumn}
\usepackage{bm} 

\newcommand{\pithKK}{\pi^+\pi^-\pi^0 K^+ K^-}
\newcommand{\pifive}{\pi^+\pi^-\pi^0\pi^+\pi^-} 
\newcommand{\pith}{\pi^+\pi^-\pi^0}
\newcommand{\ppJ}{\pi^+\pi^- J/\psi}
\newcommand{\ef}{\epsilon}
\newcommand{\effitDT}{\epsilon_{fit}^{DT}}
\newcommand{\effitMC}{\epsilon_{fit}^{MC}}
\newcommand{\efPID}{\epsilon_{PID}}
\newcommand{\efMC}{\epsilon^{MC}}
\newcommand{\efPIDMC}{\epsilon_{PID}^{MC}}
\newcommand{\efPIDDT}{\epsilon_{PID}^{DT}}
\newcommand{\psip}{\psi^{\prime}}
\newcommand{\be}{\begin{enumerate}}
\newcommand{\ee}{\end{enumerate}}
\newcommand{\bi}{\begin{itemize}}
\newcommand{\ei}{\end{itemize}}
\newcommand{\mrec}{m_{recoil}^{\pi^+\pi^-}}

\newcommand{\opi}{\omega\pi^+\pi^-}

\newcommand{\fipi}{\phi\pi^+\pi^-}

\newcommand{\fikk}{\phi K^+ K^-}
\newcommand{\ok}{\omega K^+K^-}
\newcommand{\opp}{\omega p\bar p}  
\newcommand{\fipp}{\phi p\bar p}

\newcommand{\jpsi}{J/\psi}

\newcommand{\pbar}{\overline{p}}
\newcommand{\ppbar}{p\overline{p}}

\newcommand{\rar}{\rightarrow}

\def\Journal#1&#2&#3(#4){#1{\bf #2}, #3 (#4)}

\def\NIMA{Nucl. Inst.  and Meths. {\bf A}}
\def\NPB{Nucl.  Phys.  {\bf B}}
\def\PLB{Phys.  Lett.  {\bf B}}
\def\PRL{Phys.  Rev.  Lett.  }
\def\PRD{Phys.  Rev.  {\bf D}}

\def\bec{\begin{center}}
\def\eec{\end{center}}

\begin{document}
\title{\Large \bf \boldmath Efficiency Correction in the Branching Fraction 
Measurements of $\psip$ Hadronic Decays 
\footnote{Translated from the paper which will be published in 
HEP \& NP Vol.27(2),2003 (in Chinese)}
}

\author{WANG Wen-Feng$^{1,2}$,    ZHU Yong-Sheng$^1$, ZHANG Xue-Yao$^2$.\\  
\vspace{0.2cm}
$^1$ Institute of High Energy Physics, Beijing 100039, People's Republic of
     China\\
$^{2}$ Shandong University, Jinan 250100, People's Republic of China}

\noindent\vskip 0.2cm 
\begin{abstract}
The detection efficiency correction for the particle identification and 
kinematic fit selection criteria is investigeted for the branching fraction 
measurements in the $\psip$ decays into
$\opi$, $b_1\pi$, $\omega f_2(1270)$, $\ok$, $\opp$, $\fipi$, 
$\phi f_0(980)$ , $\fikk$, $\fipp$ final states using 
 $4\times10^6 ~ \psip$ data sample collected at BEPC/BES. Based on the 
corrected efficiencies, the preliminary results of these decay channels have 
been obtained, and the "12\%" rule predicted by perturbative QCD theory 
tested.  \\ \\
\end{abstract}
\maketitle

\section{Introduction}\label{introd} 

Based on $4\times10^6 ~ \psip$ data sample \cite{4M} collected at BEijing 
Spectrometer \cite{BES}, BES collaboration measured the branching fractions
of $\psip$ decays into $\opi$, $b_1\pi$, $\omega f_2(1270)$, $\ok$, $\opp$, 
$\fipi$, $\phi f_0(980)$ , $\fikk$, $\fipp$ final states.
Perturbative QCD (PQCD) predicts \cite{PQCD} that for any
exclusive hadronic state $h$, the $\jpsi$ and $\psip$ decay branching
fractions will scale as 
\begin{eqnarray*}
Q_h=
\frac{B(\psip\rar h)}{B(\jpsi\rar h)}
\simeq\frac{B(\psip\rar e^+e^-)}{B(\jpsi\rar e^+e^-)}\simeq 12\%,
\end{eqnarray*}
where the leptonic branching fractions are taken from the PDG 
tables \cite{PDG}. This relation is known 
as the ``$12\%$ rule''. Although the rule works reasonably well  for 
a number of specific decay modes, it fails severely in the case of the 
$\psip$ two-body decays to the vector-pseudoscalar ($VP$) meson final 
states, $\rho\pi$ and $K^*\bar K$ ~\cite{rhopi}.  
This anomaly is commonly called the {\em $\rho\pi$ puzzle}.
In addition, the BES group has reported
violations of the $12\%$ rule for vector-tensor
($VT$) decay modes\cite{vt}.  Although a number
of theoretical explanations have been
proposed to explain this puzzle~\cite{puzzle}, it seems that
most of them do not provide a satisfactory solution.
Therefore, it is meaningful to measure the branching fractions of these 
two- and three- body hadronic decay channels,
and to test the  "$12\%$ rule", for the 
improvement and development of existing models. 

  The branching fraction of the decay process $\psip\rar X$ is determined by
\begin{eqnarray*}
B(\psip\rar X)=
\frac{n^{obs}(\psip\rar X\rar Y)}{N_{\psip}B(X\rar Y)\ef(\psip\rar X\rar 
Y)},
\end{eqnarray*}
where $X$ stands for the intermediate state, $Y$ the final state, $\ef$ the 
detection efficiency, $N_{\psip}$ the number of $\psip$ events, and 
$n^{obs}$ the observed signal events selected by applying certain creteria. 
The detection efficiency $\ef$ is usually determined by Monte Carlo (MC)
simulation. Take the process of $\psip\rar\ok$ as an example, the 
intermediate state $X$ is $\ok$, the final state $Y$ is $\pithKK$. We 
produce events of $\psip\rar\ok\rar\pithKK$ with phase space event generator
HOWL, which then pass through the BES detector simulation package SOBER 
\cite{BES} and we acquire the MC simulated data.  If we produce MC 
simulated data for $N_0$ signal events, reconstruct them as same as for 
the real data and apply the same selection criteria for these reconstructed 
data, finally obtain $N$ observed signal events, then, the detection 
efficiency is $\efMC$=$N/N_0$.

   If the MC simulation and the reconstruction are sufficiently accurate, 
the detection efficiency determined in this way is correspondingly accurate.
However, it is inevitable there exist much approximations in the simulation 
and reconstruction, therefore, the MC determined efficiency, $\efMC$, must 
differ from the actual value.  If the difference is relatively small, it can 
be treated as a part of systematic error.  If the deviation is relatively 
large, it is necessary to correct it to decrease the systematic uncertainty.  

\section{Deviation of $\efMC$ and Correction}
   The deviation between $\efMC$ and actual efficiency depends on the event 
selection criteria.

   The event topology for the nine channels we investigated is either four 
prong ($4P$) or four prong plus two photons produced by a neutral pion 
decay ($4P2\gamma$). The general pre-selection criteria are following:
\be
 \item The number of charged particles must be equal to four with net charge 
     zero.
 \item The number of photon candidates must be equal to or greater than two 
     for the decay channels containing $\pi^0$.  
 \item Particle identification (PID)

     For each charged track in an event, the $\chi^2_{PID}(i)$ and 
its corresponding $Prob_{PID}(i)$ values are calculated based on the 
measurements of $dE/dx$ in the MDC and the time of flight in the TOF, with 
 definitions
$$\chi^{2}_{PID}(i)=\chi^{2}_{dE/dx}(i)+\chi^{2}_{TOF}(i)$$
$$Prob_{PID}(i)=Prob(\chi^{2}_{PID}(i),ndf_{PID}),$$
where $ndf_{PID}=2$ is the number of degrees of freedom in the $\chi^{2}_{PID}(i)$
determination and $Prob_{PID}(i)$ signifies the probability of this track 
having a particle $i$ assignment. For final states containing $\ppbar$, we 
require at least one of the charged tracks satisfy 
$Prob_{PID}(p/\overline{p})>0.01>Prob_{PID}(\pi/K)$, while for other 
channels analyzed, the
probability of a charged track for a candidate particle assignment 
is required to be greater than 0.01. For instance, in the selection of 
$\psip\rar\ok\rar\pithKK$, we require that the probability of two tracks 
identified as pions must be greater than 0.01, and the probability of 
another two tracks identified as kaons also be greater than 0.01.
 \item Kinematic fit 

A 4C (4 prong events) or 5C (4 prong plus two photon events) kinematic fit 
is performed for each event. 
To be selected for any candidate final state, the event probability
given by the fit must be greater than 0.01. We will show that this 
requirement is uncorrelated to the PID requirement in our concrete event 
selection criteria (See Section V).
 \item The combined $\chi^2$

The combined $\chi^2$, $\chi_{com}^{2}$, is defined as the sum
of the $\chi^2$ values of the kinematic fit and those from
each of the four particle identification assignments: 
$$\chi_{com}^{2}=\sum_{i}\chi^{2}_{PID}(i)+\chi^{2}_{kine},$$
which corresponds to the combined probability:
$$Prob_{com}=Prob(\chi_{com}^2,ndf_{com}),$$
where $ndf_{com}$ is the corresponding total number of degrees of the
freedom in the $\chi_{com}^{2}$ determination.
The final state with the largest $Prob_{com}$ is taken as the  
candidate assignment for each event.
For example, in the selection of $\psip\rar\ok\rar\pithKK$ events, we require
$Prob_{com}(\pithKK)$ must be greater 
than $Prob_{com}(\pifive)$ and $Prob_{com}(K^+ K^-\pi^0 K^+ K^-)$
to reject possible backgrounds. 
\ee

   The efficiency of PID requirement $Prob_{PID}(i)>0.01$ is defined as
$\efPID(i)=N_i/N_{i0}$, where $N_{i0}$ and $N_{i}$ is the number of 
particle 
$i$ before and after applying the PID requirement $Prob_{PID}(i)>0.01$. 
We will show in next sections that $\efPIDMC$ differs from $\efPIDDT$.
Similary, the efficiency of kinematic fit requirement, $\effitMC$ differs 
from $\effitDT$, too. Therefore, we need to make adequate corrections for 
them. 

\section{Deviation of $\efPIDMC$ and Correction}
   If we select data samples of pure $\pi,K,p,\pbar$ particles from real 
data by a set of criteria without PID requirement, these data samples 
should be unbiased to the PID requirement. By applying the cut 
$Prob_{PID}(i)>0.01$, we obtain $\efPIDDT(i)$, which should be close to 
the real PID efficiency. The difference between $\efPIDDT(i)$ and 
$\efPIDMC(i)$ reflects the deviation of the simulated MC efficiency from 
actual value.

   We select $\pi,K,p,\pbar$ samples from processes 
$\psip\rar\ppJ$,$\jpsi\rar\pith,K^+K^-\pi^0$,$\ppbar\pi^0,\ppbar$, because
they have large branching fractions and $4P$ and $4P2\gamma$ event 
topologies, which are as same as the processes we investigated.  
Therefore, the results should be applicable for our studies. 

\bi
 \item Determination of $\efPIDDT(\pi)$

  $\efPIDDT(\pi)$ is determined with charged pion sample in
  $\psip\rar\ppJ$,$\jpsi\rar\pith$ events, which are selected by following 
  cuts:
   \be
    \item Use cuts 1,2,4 stated in Section II to select $\psip\rar\pifive$ 
               events
    \item Require $Prob_{fit}(\pifive)>Prob_{fit}(\pithKK)$ to reject possible 
               backgrounds
    \item Require $|\mrec-3.1|<0.03$ GeV to ensure the existence of 
              $\ppJ$ intermediate state, where $\mrec$ is the invariant mass 
              recoiling against the candidate $\pi^+\pi^-$ pair
   \ee

   Furthermore, we require one of daughter particles from $\jpsi$ decay 
selected with above criteria must satisfy $Prob_{PID}(\pi)>0.01$,
and $Prob_{PID}(\pi)>10Prob_{PID}(K/p)$ to ensure it is a charged pion, 
then another track must also be a charged pion, with which we form a charged 
pion sample.  This pion sample should be unbiased to pion's PID, because
it is formed without PID requirement. By applying the requirement of 
$Prob_{PID}(\pi)>0.01$ to this pion sample we can easily
determine efficiency $\efPIDDT(\pi)$.

\begin{figure}[htbp]
\centerline{\epsfig{figure=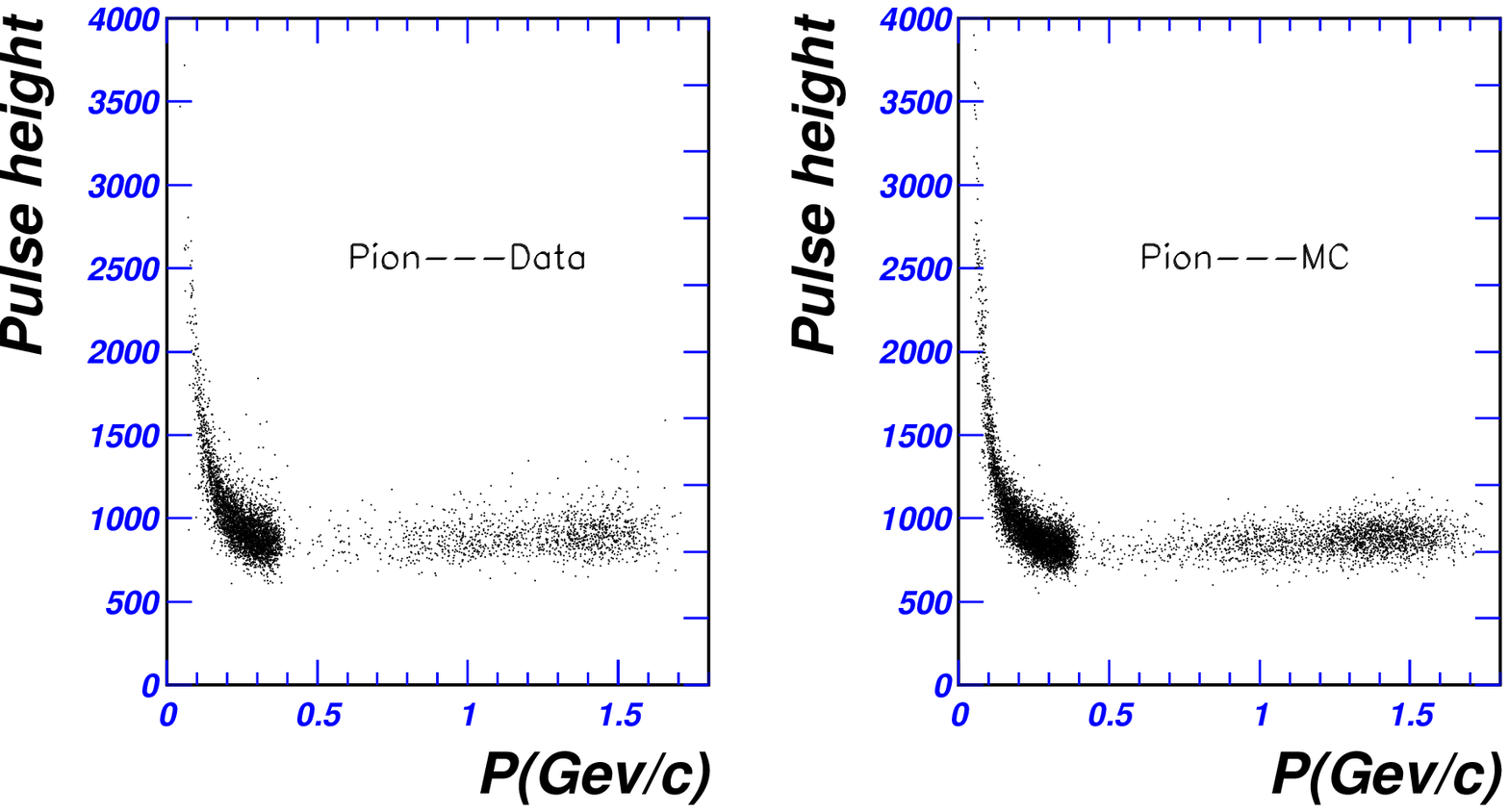,height=5.5cm,width=0.5\textwidth}}
\caption{$dE/dX$ vs. $p$ (momentum) of charged pions in
$\psip\rar\ppJ$,$\jpsi\rar\pith$ events}
\label{fig:chpion}
\end{figure}

 Fig. \ref{fig:chpion} shows the $dE/dX$ distributions of candidate charged 
pions selected with above criteria for MC and real data. In the Figure, the 
pions with higher momentum are produced from $\jpsi$ decays, while those 
with lower  momentum are the "direct" pions from $\psip\rar\ppJ$ decays. We
see that the purity of pion sample is good, no pollution from $K,p,\pbar$ 
seen. Also, the consistency between MC and real data is nice.

\item Determination of $\efPIDDT(K)$
      
  $\efPIDDT(K)$ is determined with charged kaon sample in
  $\psip\rar\ppJ$,$\jpsi\rar K^+K^-\pi^0$ events, which are selected by 
following  cuts:
   \be
    \item Use cuts 1,2,4 stated in Section 2 to select $\psip\rar\pithKK$ 
events
    \item Require $Prob_{fit}(\pifive)<Prob_{fit}(\pithKK)$ to reject possible
               backgrounds
    \item Require $|\mrec-3.1|<0.03$ GeV to ensure the existence of
              $\ppJ$ intermediate state.
   \ee
   In addition, we require one of daughter particles from $\jpsi$ decay
must satisfy $Prob_{PID}(K)>0.01$, and $Prob_{PID}(K)>10Prob_{PID}(\pi/p)$ 
to ensure it is a charged kaon, then
another track must also be a charged kaon, with which we form a charged
kaon sample.  This kaon sample should be unbiased to kaon's PID, because
it is formed without PID requirement. By applying the requirement of
$Prob_{PID}(K)>0.01$ to this kaon sample we can easily
determine efficiency $\efPIDDT(K)$.

\item Determination of $\efPIDDT(p/\pbar)$

  $\efPIDDT(p/\pbar)$ is determined with $p/\pbar$ sample in
  $\psip\rar\ppJ$,$\jpsi\rar\ppbar\pi^0$ events, which are selected by
following  cuts:
   \be
    \item Use cuts 1,2,4 stated in Section II to select 
                $\psip\rar\pith\ppbar$ events
    \item Require $Prob_{fit}(\pith\ppbar)$ must be greater then 
                $Prob_{fit}(\pifive)$ and $Prob_{fit}(\pithKK)$ 
                to reject possible  backgrounds
    \item Require $|\mrec-3.1|<0.03$ GeV to ensure the existence of
              $\ppJ$ intermediate state.
   \ee
  We further require one of daughter particles from $\jpsi$ decay
must satisfy $Prob_{PID}(p/\pbar)>0.01$, and 
$Prob_{PID}(p/\pbar)>10Prob_{PID}(\pi/K)$ to ensure it is a proton (antiproton)
, then another track must be an antiproton (proton), with which we form 
$p/\pbar$ sample.  This sample should be unbiased to ($p/\pbar$)'s PID, 
because it is formed without PID requirement. By applying the requirement 
of $Prob_{PID}(p/\pbar)>0.01$ to this sample we can easily
determine efficiency $\efPIDDT(p/\pbar)$.
\ei

\begin{figure}[htbp]
\centerline{\epsfig{figure=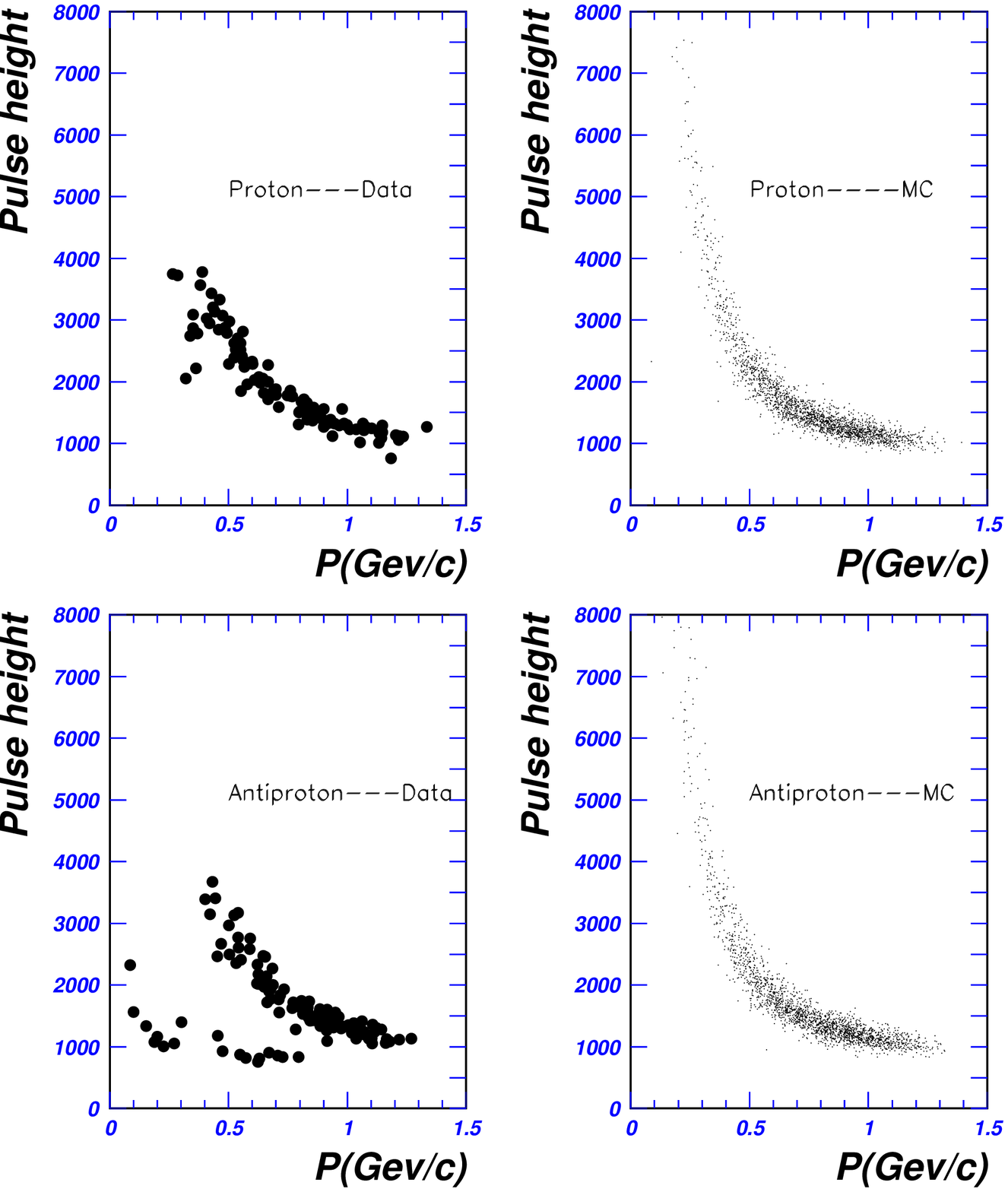,height=10.5cm,width=0.5\textwidth}}
\caption{$dE/dX$ vs. $p$ (momentum) of $p/\pbar$ in
$\psip\rar\ppJ$,$\jpsi\rar\ppbar\pi^0$ events}
\label{fig:ppbar}
\end{figure}

 Fig. \ref{fig:ppbar} shows the $dE/dX$ distributions of $p/\pbar$ samples
selected with above criteria for MC and real data. In the Figure we
see that the purity of proton sample is good, the consistency between MC and 
real data is nice; while the antiproton sample is polluted by other 
particles, which are produced by the annihilation of $\pbar$ in detector 
material and not correctly simulated in the Monte Carlo, therefore the 
MC determined efficiency has to be corrected.  With these $p/\pbar$ samples 
we determine $\efPIDDT(p/\pbar)$, and in the determination of 
$\efPIDDT(\pbar)$ only the $\pbar$ sample which $dE/dX$ values located on 
the expected band is selected to cut out other particles pollution.

Table I,II list the $\efPIDDT(i)$ and $\efPIDMC(i)$ for PID requirement 
$Prob_{PID}>0.01$, and the corresponding correction factor 
$f_{PID}(i)=\efPIDDT(i)/\efPIDMC(i)$.  From tables we know that, 
$f_{PID}(\pi)$ and $f_{PID}(K)$ are close to 1, while $f_{PID}(p)$ and 
especially $f_{PID}(\pbar)$ deviate from 1 significantly.  Therefore, for the 
final states containing $\ppbar$ pair we will not require that both 
$Prob_{PID}(p)>0.01$ and $Prob_{PID}(\pbar)>0.01$ must be 
satisfied, instead, we just 
require one of them be satisfied.  In this case the correction factor 
$f_{PID}(p/\pbar)$ is close to 1, as shown in the last row of Table II.  
Because $p$ and $\pbar$ has to be produced in a pair, if one of them is 
correctly identified, its twin is also correctly identified.

Take $\psip\rar\phi K^+K^-\rar K^+K^-K^+K^-$ as an example, if we use PID 
requirement $Prob_{PID}(K)>0.01$ to select four candidate kaons, the 
correction factor is $f_{PID}(4K)=[f_{PID}(K)]^4$.   
\vspace{0.4cm}
\begin{table}
\caption{$\efPIDDT(\pi)$,$\efPIDMC(\pi)$ and $f_{PID}(\pi)$}
\vspace{0.2cm}
\label{Table1}
\begin{ruledtabular}
\begin{tabular}{r c c c }
$p$ (GeV/c)  & $\efPIDDT(\pi)$ & $\efPIDMC(\pi)$ & $f_{PID}(\pi)$ \\\hline
$<0.5$         & $0.950\pm0.004$ & $0.966\pm0.002$ & $0.983\pm0.005$ \\
$0.5\rar 0.8$  & $0.979\pm0.022$ & $1.000\pm0.001$ & $0.979\pm0.021$ \\
$0.8\rar 1.2$  & $0.961\pm0.014$ & $0.984\pm0.009$ & $0.977\pm0.015$ \\
$1.2\rar 1.4$  & $0.956\pm0.017$ & $0.989\pm0.005$ & $0.967\pm0.017$ \\
\end{tabular}
\end{ruledtabular}
\end{table}
\vspace{0.4cm}

\begin{table}
\caption{$\efPIDDT$,$\efPIDMC$ and $f_{PID}$ for $K,p,\pbar$}
\vspace{0.2cm}
\label{Table2}
\begin{ruledtabular}
\begin{tabular}{l c c c }
Particle $i$ & $\efPIDDT(i)$   & $\efPIDMC(i)$   & $f_{PID}(i)$ \\\hline
$K^{\pm}$    & $0.952\pm0.020$ & $0.980\pm0.004$ & $0.971\pm0.020$ \\
$p$          & $0.889\pm0.035$ & $0.936\pm0.006$ & $0.950\pm0.030$ \\
$\pbar$      & $0.759\pm0.052$ & $0.929\pm0.006$ & $0.817\pm0.043$ \\
$p/\pbar$*   & $0.973\pm0.016$ & $0.996\pm0.008$ & $0.977\pm0.018$ 
\\
\end{tabular}
\end{ruledtabular}
\end{table}   
* either $p$ or $\pbar$ satisfies PID requirement
   
\section{Deviation of $\effitMC$ and Correction}
  particle reaction should conserve 4-momentum in initial and final 
states. In experiment, the measured 4-momentum of particle will 
somewhat deviate from its true values due to measurement errors.  Therefore,
the 4-momentum conservation is not accurate for measured values.  The 
kinematic fit is an algorithm, which uses least square 
priciple,  by varying measured 4-momenta of particles
to satisfy the 4-momentum conservation and other physics 
constraint (for instance, some particles in final state coming from an 
intermediate resonance decay) to the possibly best level within 
measurement errors. Kinematic fit is a powerful tool for signal events 
selection and background rejection.  In our studies
we use requirement $Prob_{fit}>a$ ($0<a<1$) to select signal events. However,
because MC could not simulate physics reality without error, this 
requirement will cause bias of $\effitMC$.

  If we select a data sample of certain events without kinematic fit, this 
sample should be bias free to the kinematic fit requirement.  By applying
$Prob_{fit}>0.01$ to this data sample, we acquire $\effitDT$, which should 
be close to the true
efficiency.  Then, the difference between $\effitDT$ and $\effitMC$ reflects 
the deviation of MC simulation from real data due to the kinematic fit 
requirement.

   We select $\psip\rar\ppJ,\jpsi\rar\ppbar$ events without using kinematic 
fit requirement by following cuts:
 \be
    \item   $p_\pi<0.6$ GeV/c (in this process, the momemtum of pions 
                  are rather low).
    \item Require $|\mrec-3.1|<0.03$ GeV to ensure the existence of
              $\ppJ$ intermediate state.
    \item require $0.9<p_{p/\pbar}<1.5$GeV/c to reject possible $\jpsi\rar 
              e^+e^-,\mu^+\mu^-$ backgrounds.
    \item Require $Prob_{PID}(p)>0.01$, $Prob_{PID}(\pbar)>0.01$, to ensure
              $\psip\rar\ppJ$,$\jpsi\rar\ppbar$ events selected.
 \ee
Based on the selected events we determine the branching fraction of 
$\jpsi\rar\ppbar$ to be $2.12\times 10^{-3}$, consistent with the 
PDG value~\cite{PDG}, which indicates the reliability of events sample.
Applying the requirement $Prob_{fit}(\pi^+\pi^-\ppbar)>0.01$ obtains 
$\effitDT(\pi^+\pi^-\ppbar)=0.773\pm 0.015$, while from MC we get 
$\effitMC(\pi^+\pi^-\ppbar)=0.948\pm 0.08$, therefore, the correction factor 
is $f_{fit}(\pi^+\pi^-\ppbar)=0.815\pm 0.017$.

Reference ~\cite{Yuan} has shown that the correction factor for kinematic 
requirement $Prob_{fit}(\pi^+\pi^-\mu^+\mu^-)>0.01$ is 
$f_{fit}(\pi^+\pi^-\mu^+\mu^-)=0.859$ in process 
$\psip\rar\ppJ$,$\jpsi\rar\mu^+\mu^-$.

Notice that, in the momentum region involved under our study, muon has the 
smallest, and $p\pbar$ the biggest interaction with the detector material in 
$\mu,\pi, K, \ppbar$ four type particles, therefore, we use same correction 
factor $f_{fit}(4P)=0.85\pm 0.05$ for all $4P$ final states in our study.

$e^+e^-\rar\gamma\gamma$ events have distinctive topology (two oppositely 
directed neutral tracks with the energy close to the beam energy, etc.), 
hence it is easy to identify and select with no need of kinematic fit 
requirement.  After the correction to the error matrix of neutral 
track~\cite{LiHB}, it is found $f_{fit}(e^+e^-\rar\gamma\gamma)=0.996$ for 
the kinematic fit requirement $Prob_{fit}(e^+e^-\rar\gamma\gamma)>0.01$.
This indicates that the existence of neutral tracks does not affect the 
correction factor of kinematic fit requirement.  Therefore, to the 
$4P2\gamma$ final states under study, we use the same correction factor
as to the $4P$ final states for kinematic fit requirement, but the 
uncertainty is increased from 0.05 to 0.08, namely, 
$f_{fit}(4P2\gamma)=0.85\pm 0.08$.

\section{Correlation between $\efPIDMC$ and $\effitMC$}
\begin{figure}[htbp]      
\centerline{\epsfig{figure=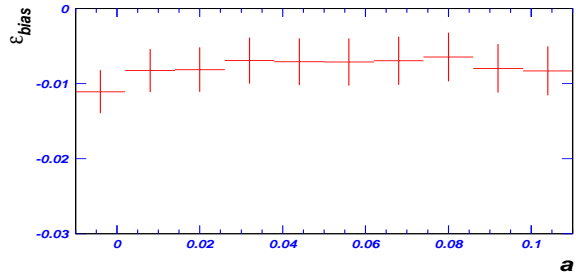,height=4.5cm,width=0.5\textwidth}}
\caption{Correlation between $\effitMC(\pi)$ and $\efPIDMC(\pi)$. See text
for the meaning of $\epsilon_{bias}$ and $a$.}
\label{fig:correl}
\end{figure}

    With the pion sample selected from $\psip\rar\ppJ$,
$\jpsi\rar\mu^+\mu^-$ events we determine $\efPIDDT(\pi)$, and using the
MC simulation for same process we determine $\efPIDMC(\pi)$.  Difine
$\epsilon_{bias}=\frac{\efPIDDT(\pi)}{\efPIDMC(\pi)}-1=f_{PID}(\pi)-1$.
Fig. \ref{fig:correl} shows the $\epsilon_{bias}$ value as the function of
$a$ for kinematic fit requirement $Prob_{fit}>a$, in which the point at   
$a<0$ denotes the kinematic fit is not applied. The statistical error is
expressed by error bar.  From the Figure we see that $\epsilon_{bias}$
(therefore $f_{PID}(\pi)$) is basically a constant for different $a$ 
within errors. This implies that the correction of PID is basically 
irrelavent to the correction of kinematic fit in our case, the correlation 
between them is rather weak, hence can be negligible.

\section{Summary}
   Based on the method described above we determine the efficiency
correction factors for nine final states listed in Table III, with
which we   give the preliminary results of  their branching fractions and
corresponding $Q_h$ values (the ratio of branching fractions of $\psip$
and $\jpsi$) ~\cite{WangWF}. Among these, the branching fractions for 
$\omega f_{2}(1270)$ and $b_{1}^{\pm}\pi^{\mp}$
supersede previous BES results ~\cite{vt,bes_ap}; while all other
branching fractions are first measurements for these decays.
The ratio of $\omega f_{2}(1270)$ (VT mode) is suppressed by a
factor of five with respect to the PQCD expectation, and
those of $\omega \pi^{+}\pi^{-}$, $\omega p\bar{p}$, and $\phi K^{+}K^{-}$
are suppressed by  a factor of two,
while those of $b_{1}^{\pm}\pi^{\mp}$ (AP mode), $\phi f_{0}(980)$ (VS   
mode), $\omega K^+K^-$ and $\phi\pi\pi$ are
consistent with PQCD expectation within errors. As to the $\phi p \bar{p}$
channel, we need more statistics.

\begin{table}
\caption{Efficiency correction factors}
\vspace{0.2cm}
\label{Table3}
\begin{ruledtabular}
\begin{tabular}{l c }
Channel   & Effi. Corr. Factor $(\%)$    \\ \hline
$\omega \pi^{+}\pi^{-}$   & $77.4\pm8.7$ \\
$b_{1}^{\pm}\pi^{\mp}$    & $77.4\pm8.7$ \\
$\omega f_{2}(1270)$      & $77.4\pm8.7$ \\
$\omega K^{+}K^{-}$       & $78.0\pm8.3$ \\
$\omega p\bar{p}$         & $85.0\pm8.0$ \\
$\phi\pi^+\pi^-$          & $75.6\pm7.6$ \\
$\phi f_0(980)(f_0\rightarrow\pi^+\pi^-)$
                          & $75.6\pm7.6$ \\
$\phi K^{+}K^{-}$         & $76.5\pm7.7$ \\
$\phi p \bar{p}$          & $85.0\pm5.0$ \\
\end{tabular}
\end{ruledtabular}
\end{table}

\acknowledgements
We acknowledge the contributions of the BEPC staff and our BES collaborators 
in the data acquisition and analyses. we are grateful to  
 Prof. Yuan Changzheng for his helpful suggestions and to Prof. Zhang Dahua 
for his work on MC simulations.
 This work  is supported in part by the National Natural Science Foundation 
of China under Contract No. 19991480, the Chinese Academy of Sciences
under contract No. KJ95T-03, and by the one-hundred talent programme 
of  CAS under Contract No. U-25.

\end{document}